\documentclass[preprintnumbers,amsmath,amssymb,pacs,12pt ]{revtex4-1}

\usepackage{graphicx}
\usepackage{epsfig}
\usepackage{amssymb}
\usepackage{graphicx}
\usepackage{dcolumn}
\usepackage{bm}
\newcommand{\ba}{\begin{array}}
\newcommand{\ea}{\end{array}}
\def\br{\begin{eqnarray}}
\def\er{\end{eqnarray}}
\def\be{\begin{equation}}
\def\ee{\end{equation}}

\def\({\left(}
\def\){\right)}

\begin{document}


\title{Chiral transition of fundamental and adjoint quarks}

\author{R.~M.~Capdevilla$^1$, A.~Doff$^2$ and A.~A.~Natale$^{1,3}$}
\affiliation{$^1$Instituto de F\'{\i}sica Te\'orica, UNESP, Rua Dr. Bento T. Ferraz, 271, Bloco II, 01140-070, S\~ao Paulo - SP, Brazil \\
$^2$Universidade Tecnol\'ogica Federal do Paran\'a - UTFPR - DAFIS, Av. Monteiro Lobato Km 04, 84016-210, Ponta Grossa - PR, Brazil \\
$^3$Centro de Ci\^encias Naturais e Humanas, Universidade Federal do ABC, 09210-170, Santo Andr\'e - SP, Brazil}

\date{\today}

\begin{abstract}

The chiral symmetry breaking transition of quarks in the fundamental and adjoint representation is studied in a model where the gap equation contains two contributions, one containing a confining propagator and another corresponding to the exchange of one-dressed dynamically massive gluons.
When quarks are in the fundamental representation the confinement effect dominates the chiral symmetry breaking while the gluon exchange
is suppressed due to the dynamical gluon mass effect in the propagator and coupling constant. In this case the chiral and deconfinement transition temperatures are
approximately the same. For quarks in the adjoint representation, due to the larger Casimir eigenvalue, the gluon exchange is operative and
the chiral transition happens at a larger temperature than the deconfinement one. 

\end{abstract}


\maketitle

Chiral symmetry breaking (csb) in non-Abelian gauge theories (NAGT), and particularly in QCD, is characterized by a non-trivial vacuum expectation value of a
fermion bilinear $\left( \left\langle {\bar{\psi}}\psi\right\rangle \right)$, by the generation of massless Goldstone bosons and
by a fermionic dynamical mass $\left( M(p^2) \right)$. This symmetry breaking has been studied for many years with the help of Schwinger-Dyson equations (SDE) and with
numerical simulations on the lattice.

In what concerns lattice simulations, it is well accepted the idea that the chiral symmetry restoration in QCD with two quark flavors in the fundamental representation is intimately connected
to the deconfinement transition \cite{bazavov,aoki}. On the other hand, when quarks are in the adjoint representation, it has been found that the chiral phase 
transition happens at a temperature ($T_c$) higher than the deconfinement temperature ($T_d$) \cite{karsch,engels,bilgici}. The ratio
between these temperatures for adjoint quarks obtained by the authors of Ref.\cite{karsch} is
\be
\frac{T_c}{T_d}\approx 7.7\pm 2.1 \, .
\label{eq0}
\ee
This result was confirmed in Ref.\cite{engels}, and a factor of order four was found in Ref.\cite{bilgici}. The comprehension of this difference is important
not only for the understanding of the csb mechanism, but it has deep phenomenological consequences for the building of technicolor
models \cite{sannino}. These results may be considered preliminary due to the fact that they were performed on coarse lattices, and larger lattices with massive
dynamical quarks may show a smooth behavior than the one described by Eq.(\ref{eq0}). Moreover, for dynamical massive quarks it seems that we even do not
have a phase transition but a crossover \cite{aoki2}, and $T_c$ should be understood as the crossover temperature. 
However, even if future improved simulations show a mild difference for these temperatures it is not evident
that they may disappear, and it is important to have information from other methods about the origin of this difference, as well as if any model indicate
a difference in these temperatures for fermions in different representations they can be ruled out by the future lattice data. 

The study of the csb mechanism in QCD through the SDE method is facing a problem due to the recent advances in their application to this theory.
The advance was the fact that the gluon and
ghost propagators studied through the SDE of pure gauge QCD \cite{aguilar} were found to be in agreement with $SU(2)$ and $SU(3)$ lattice simulations \cite{cucchieri,bogo}, indicating that the gluon may possess a dynamically generated mass, as predicted by Cornwall \cite{cornwall} many years ago. This
solution, that has been sometimes termed as ``decoupling solution", has not been frequently used in csb calculations due to an alleged gauge dependence
of the approach. The advance in this area can also be recognized after the enormous work in field theory, using the so called pinch technique, in order
to show that the SDE truncation leading to this solution is indeed gauge invariant \cite{binosi,bjc}. Recently it was also shown that the infrared
value of the dynamical gluon mass generated in the pure gauge theory is increased when the effect of dynamical quarks is added to the theory \cite{ag13,aya}. 
The problem posed to csb by the existence of a dynamically generated gluon mass for quarks in the fundamental representation is that it does not produce an amount of csb in agreement with the experimental data. This fact is clearly explained in Ref.\cite{cornwall2} and was observed in phenomenological calculations in Refs.\cite{haeri,natale0}. However, for adjoint quarks the
SDE may have an appreciable amount of csb \cite{cornwall2}.

Lattice simulations are also showing
evidences for a relation between csb and confinement, where center vortices play a fundamental role. In the $SU(2)$ case the artificial
center vortices removal also implies a recovery of the chiral symmetry \cite{lat1,lat2,lat3}, although such picture is not so clear in the $SU(3)$ case \cite{lat4}.
According to Ref.\cite{cornwall},
the SDE of NAGT have solutions that minimize the energy consistent with dynamically massive gauge bosons, leading
to an effective theory endowed with vortices, and these vortices should be responsible for confinement.  Objects like vortices cannot enter into the
SDE at the same level of ordinary Green's functions, e.g. like gauge boson propagators, since they appear in an effective theory where the quantum
effects were already taken into account, leading to dynamical gauge boson masses. 

We will discuss, in the context of the model proposed in Ref.\cite{cornwall3}, the possibility that csb, confinement and dynamical gauge boson mass generation are all intertwined in order to explain the lattice data described in the previous paragraphs. The proposal of Ref.\cite{cornwall3} consists in a fermionic SDE with two different contributions,
one generated by an effective confining propagator, whose origin can be credited to vortices, and another contribution originated by the usual exchange of $1$-dressed gluons.
The interesting point of this approach is the following:
When quarks are in the fundamental representation, the confining part of the gap equation is almost totally responsible for the quark mass
generation, while the $1$-gluon exchange barely contributes to the dynamical quark mass. When quarks are in the adjoint representation we do
have two contributions for the quark masses, one coming from the confining propagator and another coming from the $1$-dressed gluon exchange,
that has a larger contribution due to the larger Casimir eigenvalue that multiplies this particular piece of the gap equation. We expect that
exactly this extra contribution will be responsible for the difference between the chiral and deconfining temperatures for adjoint quarks,
and this is the main motivation of our work. 

If confinement is necessary and sufficient for the QCD csb we may ask how the gap equation should be modified in
order to generate a non-trivial condensate and dynamical fermion mass solution. For many years confinement has been introduced into the
gap equation in the form of the following effective confining propagator \cite{mand}:
$D^{\mu\nu}_{eff}(k) = 8 \pi K_F [{\delta^{\mu\nu}}/{k^4}]$,
whose temporal Fourier transform gives a confining linear rising potential proportional to $K_F$, which is the string tension for fermions in the fundamental representation.
This confining propagator can be associated to an area-law action with the expected confinement properties \cite{cornwall3}, but at the same
time it introduces severe infrared singularities in the fermion propagator and generates an effective Hamiltonian ($H_e$)
with only positive terms. It is impossible, with this effective Hamiltonian, to generate massless bound states, i.e. the Goldstone bosons
associated to the csb \cite{cornwall3}.

In Ref.\cite{cornwall3} it was claimed that a confining potential free of infrared singularities and still with the expected confinement 
properties is given by
\be
D_{eff}^{\mu \nu}(k) \equiv \delta^{\mu \nu} D_{eff} (k); \,\,\,\,\,  D_{eff} (k)=\frac{8\pi K_F}{(k^2+m^2)^2}   \, ,
\label{eq2}
\ee
where $m$ is a physical mass that not only cures the infrared (IR) singularities, and should be of the order of the dynamical fermion mass ($M$), but
also contributes with a negative term to the effective Hamiltonian, which is crucial to generate the massless bosons associated to the csb.
These entropic arguments, extensively discussed in Ref.\cite{cornwall3,cornwall4}, indicate that $m \sim M$ and imply in one effective 
Hamiltonian variationally minimized by the condition 
$\left\langle H_e \right\rangle =2 K_F^{1/2}-{3K_F}/{\pi M}$
in such a way that massless bound states are formed when \cite{cornwall3}
\be
M=\frac{3K_F^{1/2}}{2\pi} \, .
\label{eq4}
\ee
Of course this is a crude estimate, and it will be modified by a more detailed calculation, although it gives the correct order of magnitude value for the dynamical quark mass in the QCD case.

Taking into account the confining propagator given by Eq.(\ref{eq2}) we can write the fermionic gap equation of a general NAGT as
\br
M(p^2)&=&\int \frac{d^4k}{(2\pi)^4} D_{eff}(p-k) \frac{4M(k^2)}{k^2+M^2(k^2)} \nonumber \\
&+& C_{2R} \int \frac{d^4k}{(2\pi)^4} K(p,k)  ,
\label{eq5} 
\er
where
\[
K(p,k)=\frac{{\bar{g}}^2(k^2)3M(k^2)}{[(p-k)^2+m_g^2(p-k)][k^2+M^2(k^2)]}
\]
and $M(p^2)=M_c(p^2)+M_{g} (p^2)$ is the dynamical fermion mass generated by the confining $\left( M_c(p^2) \right)$ and one-dressed-gauge
$\left( M_{g} (p^2) \right)$ boson contributions. Note that in $M_g (p^2)$ it is assumed that the gauge boson acquires a dynamically generated mass 
$m_g (p^2)$, which also modify the effective charge to \cite{cornwall}
\be
{\bar{g}}^2(k^2)= \frac{1}{b \ln[(k^2+4m_g^2)/\Lambda^2]} \, ,
\label{eq6}
\ee
where $b=(11N-4n_fT(R))/48\pi^2$ for the $SU(N)$ group with $n_f$ flavors, and $T(R)$ is connected to the quadratic
Casimir operator ($C_2(R)$) for fermions
in one specific representation ($R$) of the gauge group. Notice that there is not double counting when we consider these two different contributions
to the gap equation. The confining propagator results from the vortices which appear in a massive effective theory after the quantum corrections
where taken into account for the gluonic Green's functions.  

The $M_{g} (p^2)$ part of the gap equation has its kernel
damped by the finite IR behavior of the gauge boson propagator, as well as the finite behavior of the coupling at $k^2=0$. This
scenario of a dynamically generated gauge boson mass has been shown to be consistent with lattice simulations \cite{cucchieri,aguilar}. The soft IR behavior 
of the kernel in the $M_g (p^2)$ part of the gap equation is what attenuates the csb for fermions in the fundamental
representation \cite{cornwall2,cornwall3}. The attenuation is not so strong for fermions in the adjoint representation
due to the larger Casimir eigenvalue of this representation. In the QCD case the $m_g$ phenomenologically preferred value 
is $m_g \approx 2\Lambda $ \cite{cornwall,natale}, where $\Lambda$ is the characteristic QCD scale. In order to perform our calculation we shall need the value of the gluon mass in the case of QCD with adjoint quarks, which appear in the $M_g(p^2)$ part of the gap equation given by Eq.(4). Unfortunately, up to now the value of the gluon mass has only been calculated in the case of pure gauge QCD \cite{aguilar, cucchieri}, and in preliminary
SDE\cite{ag13}  and lattice calculations \cite{aya}  taking into account dynamical quarks in the fundamental representation. Therefore we will assume
the gluon mass in the case of QCD with dynamical adjoint quarks as a free parameter.

We are working in the rainbow approximation and it is possible that for more sophisticated vertex the results may be changed by a few percent, but,
as observed in Refs.\cite{cornwall3,dfn}, in this approximation the results for the chiral symmetry breaking parameters in the
case of quarks in the fundamental representation are already satisfactory, and for the adjoint representation the uncertainties in
the string tension and dynamical gluon mass values may overwhelm the possible gain with an improved vertex.   

Assuming $K_F=0.21\rm{GeV}^2$ and $m=0.9M$ we can compute the dynamical quark mass from the gap equation only with the confining propagator, finding for 
fundamental quarks $M_F=M_F(0)=212$MeV, what is in agreement with the results of Refs.\cite{cornwall3,dfn}. If we now assume the breakable adjoint string tension of
$SU(N)$ NAGT given by $K_A \approx 2K_F$, which is a better approximation the larger is the $N$ value, we obtain for adjoint quarks $M_A=M_A(0)=300$MeV.
It is clear that the actual computation of quark masses and condensates should involve the complete gap equation.
We have computed the infrared mass value for fundamental quarks from Eq.(\ref{eq5}) assuming $\Lambda = 300$MeV, $m_g/\Lambda = 2$
and $m=0.9 M$. We obtained $M_F=M_F(0)= 221$MeV, meaning that the one-gauge boson exchange contributes less than $10\%$ to the total mass!
As a consequence this infrared value is basically the same if we assume $2$ or $6$ quarks, because the main dependence on
the number of flavors is contained in the one-gluon exchange contribution. The csb of fundamental quarks is totally dominated
by the confinement part of the gap equation.

\begin{figure}[ht]
\centering
\includegraphics[width=0.9\columnwidth]{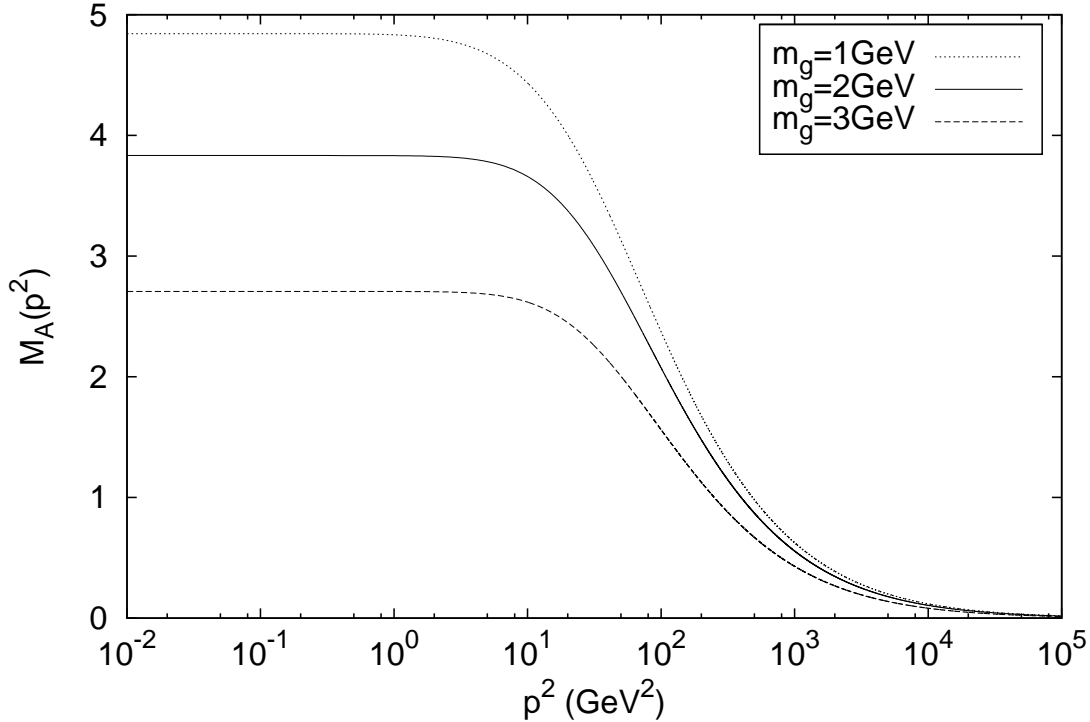}
\caption[dummy0]{Dynamical masses for adjoint quarks computed with Eq.(\ref{eq5}) and different dynamical gluon masses.} 
\label{figcomp}
\end{figure}

The dynamical masses for adjoint quarks are shown in Fig.(\ref{figcomp}). These masses were computed with $K_A \approx 2K_F = 0.42\rm{GeV}^2$, $m=M(0)$, $n_f=2$, $\Lambda = 300$MeV,
and different dynamical gluon masses ($m_g =1,2,3 \,$GeV). The reason for the different $m_g$ values is that adjoint quarks screen the gluon exchange.
As this force is proportional to the invariant product of the coupling constant times the gluon propagator the only way to lower the strength is increasing
the dynamical gluon mass. This increase of the dynamical gluon mass is already observed when fundamental dynamical quarks are added to the pure gauge theory \cite{ag13,aya}, and we should expect further increase of the dynamical gluon masses in the case of QCD with adjoint quarks. Unfortunately there is no lattice data for gluon masses associated to adjoint quarks.
Comparing the masses of Fig.(\ref{figcomp}) to the result of the adjoint quark mass discussed in the previous paragraph, obtained with the confining propagator, we verify that the effect of the confining
propagator is just the addition of a small mass to the adjoint quarks. 
We also calculated the same masses varying the factor $m$ in the confining propagator (Eq.(\ref{eq2})), and found that this factor does not change the result,
because the symmetry breaking for adjoint quarks is basically dominated by the gluon exchange. 

The other parameter that characterize the chiral transition is the fermion condensate, which is given by
\be
\left\langle {\bar{\psi}}\psi \right\rangle_R (\kappa ^2) = -\frac{N_R}{4\pi^2}\int_0^{\kappa^2} dp^2 \frac{p^2M_R(p^2)}{[p^2+M_R^2(p^2)]} \, ,
\label{eq7}
\ee
where the fermions are in a given representation $R$ with dimension $N_R$, and $M_R(p^2)$ is its dynamical mass. Note that the condensate
is computed at the scale $\kappa^2$, where it can be compared to the experimental value, and, more important, we are computing the condensate for
massless quarks, or in the chiral limit where Eq.(\ref{eq7}) does not need renormalization.
In QCD with fundamental quarks the phenomenological value of the condensate is known to be $\left\langle {\bar{q}}q \right\rangle_F (1$ GeV$^2)=(229\pm 9$MeV$)^3$ \cite{gi}.
However, we do not know what scale we have to choose in order to characterize the condensates or chiral transition for the adjoint quarks. We may expect that the
adjoint condensates are saturated at one specific scale, and we propose that this scale is given by $\kappa = 3M$, because at this scale of momentum the value of the dynamical mass has
already dropped to half of its infrared value. For instance, Eq.(\ref{eq7}) can be crudely estimated assuming $M_R(p^2)=M$ constant in the infrared region resulting in
\be
\left\langle {\bar{\psi}}\psi \right\rangle_R (\kappa^2) \approx -  \frac{N_R M^3}{4\pi^2}  \left[ \frac{\kappa^2}{M^2}-\ln \left(1+\frac{\kappa^2}{M^2}\right)\right]\, .
\label{eq7b}
\ee
Let us now assume $\kappa = 3M$ for QCD with fundamental quarks, which gives
$\left\langle {\bar{\psi}}\psi \right\rangle_R ([3M]^2) \approx - {6.7 (N_R M^3)}/{(4\pi^2)}$.
At this scale we obtain $\left\langle {\bar{\psi}}\psi \right\rangle^{1/3}_F \approx 180$MeV.
As a simple leading order calculation we see that this approximation gives the right order of magnitude for the condensate, and it will
be used to compute the adjoint condensates for the solutions presented in Fig.(\ref{figcomp}).
The condensates associated
to the infrared adjoint quark masses $M_A = 4.8,3.8,2.7 \, $GeV are: $\left\langle {\bar{\psi}}\psi \right\rangle_A^{1/3}=5.3,4.2,3.0 \, $GeV respectively, and were computed with Eq.(\ref{eq7b}) assuming
$\kappa= 3M$, which, according to Fig.(\ref{figcomp}), is also the order of the scale where the adjoint masses start decreasing.

The fermion condensate at finite temperature ($T$), following the real-time formalism of Dolan and Jackiw \cite{dj}, is given by 
\br
& &\left\langle {\bar{\psi}}\psi (T) \right\rangle 
= \left\langle {\bar{\psi}}\psi (0) \right\rangle (\kappa^2) + N_R \int \frac{d^4k}{(2\pi )^4}\times \nonumber \\
 && Tr\left\{ \frac{({\not\! k}+M(k))2\pi\delta(k^2-M^2(T))}{e^{\left|k_0/T\right|}+1}\right\} ,
\label{eq16}
\er
where $\left\langle {\bar{\psi}}\psi (0) \right\rangle (\kappa^2)$ is the zero temperature condensate calculated at the scale $\kappa^2$, which
is going to be assumed as the same one used to compute the condensate in Eq.(\ref{eq7b}). $M(T)$ in Eq.(\ref{eq16}) is proportional to $M$, the
zero temperature mass, plus a function that depends on $T$ and all other parameters appearing in the fermionic gap equation (\ref{eq5}).

The temperature dependent integral on the right hand side of Eq.(\ref{eq16}) can be easily determined and we obtain
\be
\left\langle {\bar{\psi}}\psi (T) \right\rangle = \left\langle {\bar{\psi}}\psi (0) \right\rangle (\kappa^2) + \frac{2N_R M}{\pi^2} 
T^2 J\left(\frac{M(T)}{T}\right) \, ,
\label{eq17}
\ee
where 
\be
J(\Delta)=\int_0^\infty \frac{dy \,\, y^2}{(y^2+\Delta^2)^{1/2}} \left[ e^{(y^2+\Delta^2)^{1/2}}+1\right]^{-1} \, .
\label{eq17a}
\ee

There are two main quantities that we need to know to compute the condensate at finite temperature. One is its value at zero temperature and the
other is the functional expression for $M(T)$. In the case of fundamental quarks all quantities that enter in the $T=0$ condensate calculation
are reasonably known, and $M(T)$ may be assumed to be a constant and approximately equal to its zero temperature value as long as we remain
below the critical temperature region. Of course, within this approximation
that has already been used in the literature \cite{scadron}, we may estimate the order of $T_c$, although we will know nothing about the type of the transition,
which will depend strongly on the dynamics, or the $T$ behavior of the dynamical mass in the crossover neighborhood. In the case of adjoint quarks
we can only roughly estimate the factor $m$ in Eq.(\ref{eq2}) \cite{us}, but this is totally irrelevant because most of the chiral breaking is generated
by the one-dressed gluon exchange, and we will leave the results as a function of the dynamical gluon mass, because there are not determinations
of this quantity in the literature. In this case we can also disregard the temperature dependence on $M(T)$, because of the flat infrared behavior 
observed in Fig.(\ref{figcomp}), and remember the the result that we shall obtain in the sequence is not valid at high temperatures, but will just
serve to determine the order of $T_c$.

The critical temperature ($T_c$) characterizing the phase transition is obtained from the equality
$\left\langle {\bar{\psi}}\psi (T_c) \right\rangle = M(T_c)=0 $,
which, with $J(0)=\pi^2/12$, gives the following result
\be
T_c^2= -\frac{6\left\langle {\bar{\psi}}\psi (0) \right\rangle}{N_R M}
\label{eq18a}
\ee

To obtain the deconfinement temperature we can use the $1/d$ expansion result of Pisarski and Alvarez \cite{pia} valid for a $d$ dimensional
NAGT
\be
T_d^2 = \frac{3}{\pi(d-2)} K_R  \, .
\label{eq19}
\ee
This approximation is crude for QCD but gives the correct order of magnitude of the deconfinement temperature. 
Therefore, our final result is:
\be
\left( \frac{T_c}{T_d} \right)^2\approx \frac{6.7}{\pi} \frac{M^2}{K_R} \, ,
\label{eq20}
\ee
and the ratio of temperatures are shown in Table I, which are of the order of the ratio found in lattice simulations. 

\vspace{1.5cm}
\begin{table}
\begin{center}
 \begin{tabular}{|c||c|c|c|c|c|}
 \hline 
 \vspace{0.05cm}
 Representation & $m_g$ (GeV)  &  $ M(0)$ (GeV) & $T_c/T_d$ \tabularnewline
 \hline 
 \hline
 fundamental & 0.6  & 0.22 & 0.76  \tabularnewline
 \hline
 adjoint & 1.0  & 4.8 &  10.9 \tabularnewline
 \hline 
 adjoint & 2.0   & 3.8 & 8.56 \tabularnewline
 \hline 
 adjoint & 3.0  & 2.7 & 6.08 \tabularnewline
 \hline 
 \end{tabular}
 \caption{Dynamical masses for fundamental and adjoint quarks and the ratio between the chiral and deconfinement transition temperatures given by Eq.(\ref{eq20}).}
 \label{tableT}
\end{center}
 \end{table}

We conclude saying that the Cornwall's model of csb \cite{cornwall3} lead to differences in the ratio
of the chiral to the deconfinement transition temperatures for fundamental and adjoint quarks. The fundamental chiral transition is a result of
confinement produced by vortices, whereas the adjoint chiral transition is basically driven by the dynamically massive one-gluon exchange.
Our calculation contains several approximations and certainly may be improved with new lattice determinations of the dynamical gluon masses
for QCD with adjoint quarks, although it is already surprising that the simple analysis
shown here can provide a picture of the $T_c/T_d$ ratio for fundamental and adjoint quarks that is similar to the preliminary lattice calculations
of this ratio.


\par 

\section*{Acknowledgments}
We are indebted to A. C. Aguilar for discussions and help with the numerical calculation.
This research was  partially supported by the Conselho Nacional de Desenvolvimento Cient\'{\i}fico e Tecnol\'ogico (CNPq) and by
Coordena\c c\~ao de Aperfei\c coamento de Pessoal de N\'{\i}vel Superior (CAPES).

\end{document}